# Maximum brightness theorem for waves

David A. B. Miller

*Ginzton Laboratory, Stanford University, 348 Via Pueblo Mall, Stanford, CA 94305, USA*
dabm@stanford.edu

We introduce a universal bound on the separable powers of a wave field after passing through arbitrary passive optical or wave systems. Any wave field can be expressed as a combination of mutually incoherent and mutually orthogonal components, each with some power or "brightness". We prove that, in passing through a lossless or lossy optical system, writing the components in order of power, the power in each such component at the output cannot exceed the power in each such component at the input, even though each resulting output may be an arbitrary mixture of the inputs. This result encompasses previous brightness theorems, has several immediate consequences, and gives a simple limit to the concentration of light, radio-frequency or other waves into single-mode outputs.

We often need to know with waves just how their power can be concentrated. For a single-frequency coherent wave from a laser beam or radio-frequency (r.f.) antenna, diffraction and wave propagation theory give simple exact answers. But much wave radiation, especially in the natural environment or from thermal light sources or light-emitting diodes, is only partially coherent – a sum or superposition of emissions from multiple different sources with no particular phase relations. Classical optics needed answers for applications such as solar energy concentration and conversion efficiency and in projection and displays (see (*1*) and (*2*) for recent discussions). The Constant Brightness (or Radiance) Theorem (*3*), an important basic result in classical optics, proposed that power per unit area per unit solid angle (the radiance or brightness) is conserved in loss-less systems such as with lenses and mirrors or even non-imaging concentrators (*4*) (and such brightness cannot increase in lossy systems (*4*)). It is a powerful result that works well for systems much larger than a wavelength, with incoherent emission, and with light intensities that are locally relatively uniform. However, it is based on a ray picture of light, which does not include diffraction, so it is of limited use for nanophotonic structures or r.f. antennas. Much modern discussion of optics and r.f. waves is in terms of countable modes or channels that cannot be described with rays (*5*). We want a clear and simple theorem that covers existing results but applies broadly to these modern topics.

We introduce a Maximum Brightness Theorem that applies universally across classical and modern pictures, including quantum mechanical descriptions. It encompasses previous results for coherent and partially coherent waves, building on the body of work on partial coherence (*6*) and in particular extending important recent results and approaches (*1*, *2*). It applies across size scales from sub-wavelength scatterers or antennas to astronomical optics, to electromagnetic waves of all frequencies, and to other waves more generally. This Theorem can be stated simply and exposes clear bounds for such waves in any passive linear wave system (i.e., one without gain).

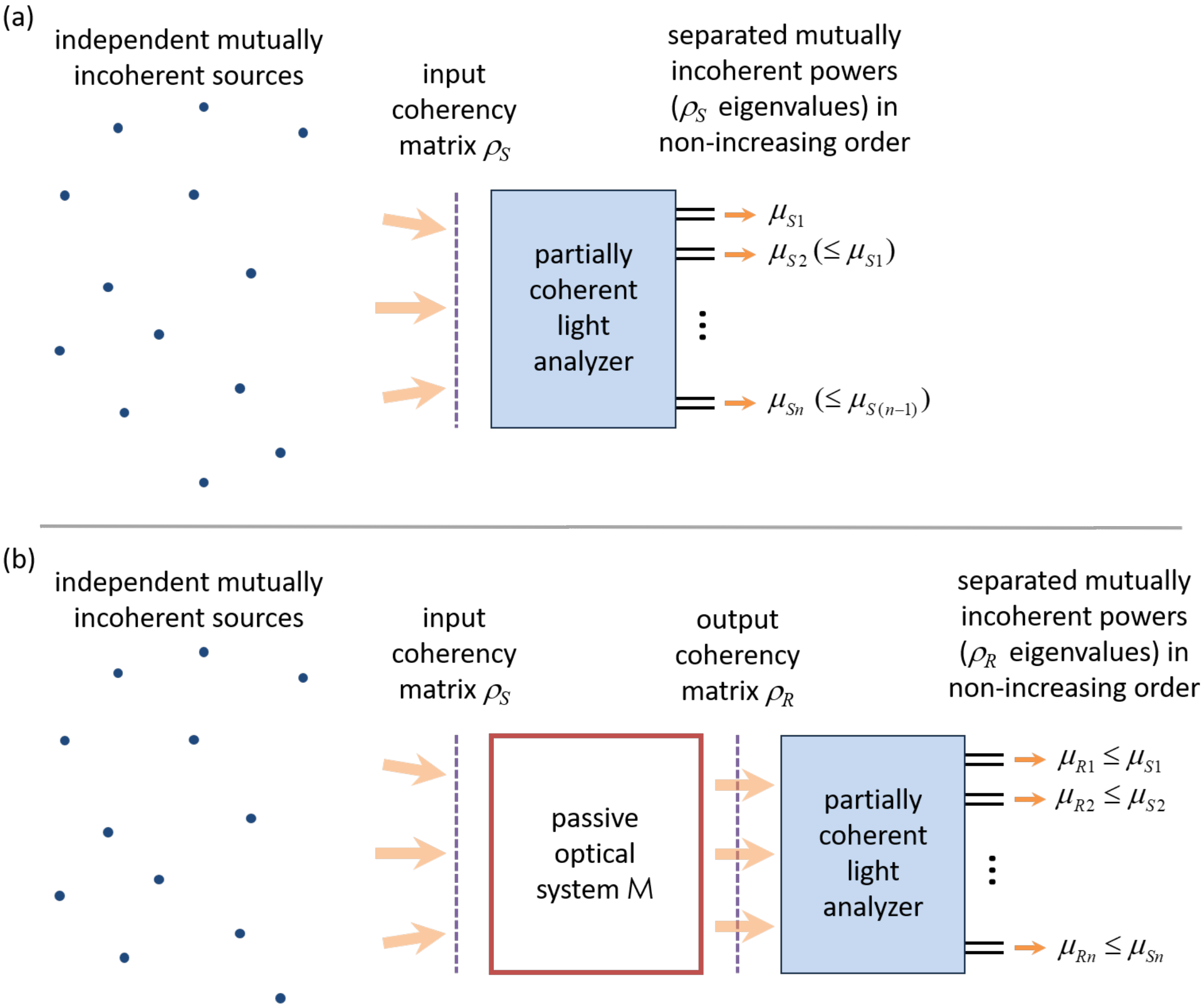

Fig. 1. Apparatus to separate partially coherent fields. (a) Illustration of light from a set of mutually incoherent sources, such as atoms emitting spontaneously, that arrives at some surface, where it can be described by an input coherency matrix $\rho_S$. That light can be separated by a (lossless) partially coherent light analyzer (PCLA) (*7*, *8*) to separate out (input) powers $\mu_{Si}$, which are the eigenvalues of $\rho_S$, in order from most powerful downwards. (b) A passive optical system M generally mixes the input field, possibly with loss, to generate an output field described by coherency matrix $\rho_R$. The Maximum Brightness Theorem states that the resulting mutually incoherent output powers $\mu_{Ri}$ (which can also be physically separated by a PCLA), in order from most powerful downwards, cannot exceed the corresponding input powers, i.e., $\mu_{Ri} \leq \mu_{Si}$, even though the system M may have mixed all the input powers into each output power.

The idea of this Theorem is shown in Fig. 1. We imagine a set of sources – e.g., atoms – each emitting independently with no specific relative phase relations. These sources are mutually incoherent – no interference pattern would emerge from superposing their resulting waves. When this radiation arrives at some optical system, generally it can be described using some finite number $n$ of orthogonal functions or modes, which may be much smaller than the number of original incoherent sources; e.g., though the sun has some very large number of atomic emitters, our eye on its own would only resolve ~1000 such modes. As a result, this input radiation is generally partially coherent, conveniently described mathematically by an input or source coherency matrix $\rho_S$ and, equivalently, by a set of orthogonal functions or components – the coherent modes – that are mutually incoherent (*9*). Recently, a partially coherent light analyzer (PCLA) (*7*, *8*) has been proposed and demonstrated that can separate such components physically into a set of single-mode outputs – e.g., as in Fig. 1(a) – concentrating their power optimally.

Importantly, the PCLA performs this separation in order, routing the most powerful such component to the first output, with power $\mu_{S1}$, the second most powerful (power $\mu_{S2}$) to the second output, and so on, in decreasing (or, more correctly, non-increasing) order. Technically, these powers are the eigenvalues $\mu_{Si}$ of $\rho_S$, with corresponding eigenvectors $|\eta_{Si}\rangle$; those eigenvectors are the coherent modes of this partially coherent wave field. (Note that the PCLA does not change the eigenvalues; it merely conveniently separates out the corresponding powers; it also measures these coherent mode functions $|\eta_{Si}\rangle$ (*7*, *8*)). Incidentally, if we were to try to interfere any two of these outputs from the PCLA, we would find no interference fringes, as recently demonstrated (*8*); these components are both mutually orthogonal and mutually incoherent.

Now we pass the input radiation through an arbitrary passive linear optical or wave system, described by linear operator $\mathsf{M}$, as in Fig. 1(b). The result will be an output or "receiving" coherency matrix $\rho_R$.We could imagine similarly analyzing and separating this with a PCLA, leading to output powers $\mu_{Ri}$ (the eigenvalues of $\rho_R$), similarly in non-increasing order (with corresponding eigenvectors and coherent modes $|\eta_{Ri}\rangle$). The key result of this paper, and its most compact statement, is that we are guaranteed that

$$\mu_{Ri} \le \mu_{Si} \tag{1}$$

for every one of these outputs (or the underlying components), no matter what the passive optical or wave system $\mathsf{M}$ is. We can think of powers $\mu_{Si}$ or $\mu_{Ri}$ as each being a modal "brightness", so equivalently we can state our theorem:

**Maximum Brightness Theorem**: In passing through a passive optical system, the power in the *i*th most powerful mutually incoherent mutually orthogonal component of the field cannot increase (2)

This theorem sets simple and universal bounds on our ability to concentrate wave radiation by any linear passive wave system, with or without loss.

Note the separation of these powers by the PCLA is just for our conceptual convenience. The mutually orthogonal mutually incoherent components still exist, and the Theorem and its expression in Eq. (1) are still true even without this explicit separation by the PCLA.

One might imagine this Theorem is obvious because we are just establishing some set of channels through the system and the power is conserved or attenuated on each such channel, but this is not generally the case. The *i*th such output component can be made up from any arbitrary combination of the corresponding input components, as mixed by the passive optical system. If we had modulated the input components individually with some signals, each output component would in general have a mixture of all these signals. The Theorem says that, despite any such arbitrary mixing and possible loss, we still cannot increase the power in the *i*th most powerful component.

## Background to the proof

The proof of this Theorem rests on two key steps. First, the communication mode description of wave systems (*5*, *10*) shows that there are discrete and countable orthogonal (i.e., zero cross-talk) wave channels through any linear optics or scatterer (*11*), mapping orthogonal sources or waves in one "source" volume or surface one by one to orthogonal waves in another "receiving" volume or surface. Such a modal picture has seen increasing applications and results in optics, e.g., (*12–14*), and r.f. waves, e.g., (*15–18*). It also shows that only a finite number of such modes is required to describe any wave that can be practically communicated. This finiteness is protected in several ways: first, there is an overall sum rule on "power" coupling strengths (*5*, *10*); second, empirical calculations, e.g., (*5*, *13*, *17*, *19*, *20*), on specific source and receiving spaces always show a practical cut-off in coupling strengths after some number of communication modes; third, this practical cut-off, and the quasi-exponential fall-off in coupling strength beyond it, has recently been explained physically as the unavoidable onset of a tunneling escape of waves (*19*). So, beyond some number *n*, which we could choose at the cut-off or even as far past it as we like, subsequent channels can be truly negligible. Hence, we can use finite matrices to describe our wave systems. To be clear, these communication mode channels do *not* in general correspond to mapping specific input coherent modes $|\eta_{Si}\rangle$ to the corresponding output coherent modes $|\eta_{Ri}\rangle$, but these results show the required basis sets in input and output space are practically finite. (The ability to use finite *n* is technically important in the use of Weyl's monotonicity theorem below.)

A second step is to exploit the coherency matrix $\rho$ as a convenient way to describe multimode partially coherent light. Already well known for describing partially polarized light (*6*) and extended to include spatial modes (*21*), a major advance was to exploit this for arbitrary spatial multimode states and deduce important properties and theorems for lossless optical systems (*1*, *2*), including concepts for brightness. Here we concentrate just on spatial coherence; we presume quasi-monochromatic radiation (*6*) – i.e., sufficiently narrow bandwidths that the coherence time exceeds any difference in propagation times through the system – or at least we divide the analysis into such narrow bands.

Partially coherent light has been extensively analyzed theoretically (*6*, *9*), with one useful measure for quasi-monochromatic spatially-partially-coherent light being the mutual intensity (*6*, *9*) $J(\mathbf{r}_1, \mathbf{r}_2)$ between light at two points $\mathbf{r}_1$ and $\mathbf{r}_2$; the coherency matrix can be regarded as the representation of this function on a modal rather than position basis. In this coherency matrix picture, we start by presuming *q* different (and mutually incoherent) sources with powers $P_i$ (e.g., like those in Fig. 1), each of which generates some wave of the (normalized) form $|\chi_i\rangle$; these waves $|\chi_i\rangle$ need not be orthogonal and generally will not be. Though the waves from different sources are mutually incoherent, each such wave is coherent with itself; each $|\chi_i\rangle$ is a quite definite function, not a random variable – it might, for example, be a spherical wave emerging from a point source such as an atom.

Here, we use a generic Dirac "bra-ket" notation in which $|\chi\rangle$ can be regarded as a column vector of numbers that represent the field ($\langle\chi|$ is the corresponding row vector with complex conjugated elements), and the superscript "†" indicates the Hermitian adjoint (conjugate transpose). In a position basis, those numbers could just be the amplitudes of the field at some suitable dense set of points. In a decomposition on a modal basis, those would be the amplitudes of the various modal functions. Note that we can be considering scalar waves or any kind of vector waves, including all polarizations; those different kinds of waves just have different basis functions as necessary to represent their properties (e.g., different polarizations).

The coherency matrix for such a field is defined as

$$\rho = \sum_{i=1}^{q} P_i |\chi_i\rangle\langle\chi_i| \tag{3}$$

This form is very similar to the quantum-mechanical density matrix. Indeed, we can interpret $P_i$ as being a probability, e.g., per unit time, that a photon came from source *i*, allowing us to relate to quantum descriptions of the field. One distinction compared to usual density matrices is that we do not renormalize the coherency matrices (e.g., to unit total power or unit trace) as we proceed through the system, because it is precisely the loss in that power, as expressed through the possible reduction in eigenvalues, that underlies Eq. (1).

$\rho$ is Hermitian, so it has real eigenvalues $\mu_i$ with orthogonal eigenvectors $|\eta_i\rangle$, and is also positive semi-definite; i.e., for any $|\chi\rangle$ in the space of interest in the space of interest

$$\langle\chi|\rho|\chi\rangle \geq 0 \tag{4}$$

so $\mu_i \geq 0$ for all *i*.

These eigenvectors $|\eta_i\rangle$ of the coherency matrix can be called the coherent mode (*9*) basis. Until recently, their existence was primarily of mathematical interest because we had apparently no good way of separating a given partially coherent field onto this basis. Recently, a mesh of interferometers functioning as a partially-coherent light analyzer (PCLA) (*7*, *8*), configured variationally by maximizing output powers, has physically separated partially coherent light into these modes, with the resulting powers in the output waveguides being the corresponding eigenvalues $\mu_i$, arranged in non-increasing order. (The eigenvectors $|\eta_i\rangle$ can be deduced directly from the resulting settings of the interferometers in the mesh.) An alternative approach measures the coherency matrix tomographically (*22*, *23*), allowing the coherency matrix to be calculated; if an interferometer mesh is calibrated and then programmed accordingly (implementing the same matrix as in the PCLA), it similarly separates the field into its coherent mode components.

Now consider an arbitrary passive linear optical system that takes an input (coherent) wave $|\chi\rangle$ in our *n* dimensional space and linearly transforms it through some $n \times n$ matrix $\mathsf{M}$ to generate an output wave $|\zeta\rangle$, i.e.,

$$|\zeta\rangle = \mathsf{M}|\chi\rangle \tag{5}$$

Then, for some input or "source" coherency matrix $\rho_S$, the corresponding output or "received" coherency matrix is

$$\rho_R = \mathsf{M}\rho_S\mathsf{M}^\dagger \tag{6}$$

This relation holds even if $\mathsf{M}$ represents a lossy system. (For completeness, we give the proof in Supplementary text section S1.)

As final background, consider first the case of a lossless optical system, which can be represented by a unitary operator $\mathsf{U}$, and some partially coherent "input" field represented by the coherency matrix $\rho_S$, written on its coherent mode eigenfunction basis $|\eta_{Si}\rangle$ as

$$\rho_S = \sum_i \mu_{Si} |\eta_{Si}\rangle\langle\eta_{Si}| \tag{7}$$

From Eq. (6), the unitary optical system transforms the coherency matrix to give a new or "output" coherency matrix

$$\rho_R = \mathsf{U}\rho_S\mathsf{U}^\dagger \tag{8}$$

Using Eq. (7), we can write this as

$$\rho_R = \mathsf{U}\left(\sum_i \mu_{Si} |\eta_{Si}\rangle\langle\eta_{Si}|\right)\mathsf{U}^\dagger = \sum_i \mu_{Si}\, \mathsf{U}|\eta_{Si}\rangle\langle\eta_{Si}|\mathsf{U}^\dagger \tag{9}$$

so, writing

$$|\eta_{Ri}\rangle = \mathsf{U}|\eta_{Si}\rangle \tag{10}$$

which also form an orthonormal set because the unitary transform preserves orthogonality and "length", we have

$$\rho_R = \sum_i \mu_{Si} |\eta_{Ri}\rangle\langle\eta_{Ri}| \tag{11}$$

So, under any such unitary transformation (such as with the PCLA), the coherency matrix transformed through a lossless system has the same eigenvalues (and hence powers) as the original coherency matrix (see, e.g., recent discussions (*1*, *2*, *24*)). Note this is true even though the output modes are in general combinations of the input modes (as given by Eq. (10)). This conservation of eigenvalues in the lossless case means also that, if the rank (the number of non-zero eigenvalues) of the coherency matrix is finite, then it is conserved (*2*) in a lossless system.

If there is loss that is uniform over all the modes of interest, at some multiplying amplitude loss factor $\gamma$, then we can simply replace $\mathsf{U}$ by $\gamma\mathsf{U}$ above and follow through similar algebra to deduce that the eigenvalues of the output density matrix will become just scaled versions of the input eigenvalues, i.e., $|\gamma|^2 \mu_i$. Even though the eigenvalues are not conserved, their relative magnitudes are still maintained under such uniform loss, and the rank of any finite-rank coherency matrix is also conserved.

## Proof of the maximum brightness theorem

For the general case of loss that is not necessarily the same for all channels, the situation is more complicated. Ref. (*1*) shows that the largest eigenvalue cannot increase in passing through some passive system. Ref. (*2*) goes further to show that the input eigenvalues majorize the output eigenvalues (see Supplementary text S5). Here we prove the tighter bound as in Eq. (1). We give the full mathematical proof in the Methods section, but we can summarize it here.

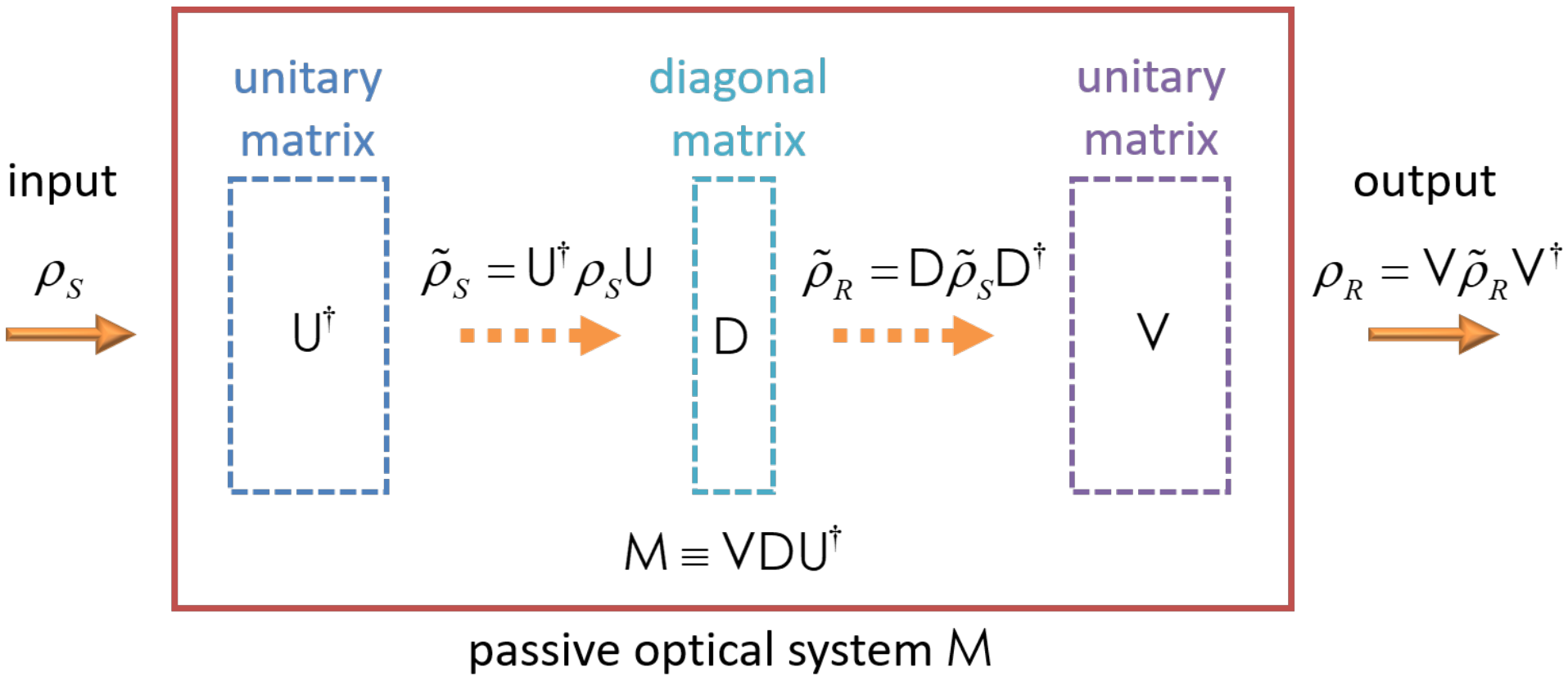

Fig. 2. Conceptual view of transforming a partially coherent light field. An input light field represented by coherency matrix $\rho_S$ is changed by a passive optical system, represented by matrix $\mathsf{M}$, to generate an output field represented by coherency matrix $\rho_R$. The mathematical sequence of operations based on the singular-value decomposition (SVD) of $\mathsf{M}$ is also indicated, with corresponding intermediate coherency matrices. If $\mathsf{M}$ was implemented using a corresponding sequence of optical components (*25*), these intermediate coherency matrices would also physically exist.

Briefly, we consider an arbitrary passive optical system $\mathsf{M}$ as in Fig. 2, describing the $n \times n$ matrix $\mathsf{M}$ through its singular-value decomposition (SVD) $\mathsf{M} = \mathsf{V}\mathsf{D}\mathsf{U}^\dagger$ where $\mathsf{U}$ and $\mathsf{V}$ are unitary and $\mathsf{D}$ is a diagonal matrix of singular values $s_i$, with $|s_i|^2 \le 1$ because the system has no gain. Unitary transformations preserve the eigenvalues (as discussed above), so we can be comparing the eigenvalues of the transformed coherency matrices $\tilde{\rho}_S$ and $\tilde{\rho}_R$. By mathematical manipulations on these matrices and their eigenvalues, involving standard results of the square roots of positive definite matrices, the cyclic property of matrix products, and especially the Weyl monotonicity theorem (with proofs given in the Supplementary text sections S1 to S4), we can prove the result Eq. (1).

## Discussion and extensions

This theorem has some immediate extensions and applications.

First, we note that, though presented here as transmission from an input on the left to an output on the right, the results are valid for the more general scattering matrix approach, and for non-reciprocal systems (Supplementary text S6).

Second, for any passive optical system M, if we know its $|s_j|^2$ (for the singular values $s_i$), we can deduce the maximum brightness (power) of any mutually orthogonal, mutually incoherent component of the output, as well as the maximum total power through the system for any partially coherent input field (Supplementary text S7).

Third, if we define a "threshold rank" $r_\varepsilon$ as the number of the source powers $\mu_{Si} > \varepsilon$ for some threshold power $\varepsilon$, then we can prove $r_\varepsilon$ cannot increase in any passive optical system (Supplementary text section S8). This may be useful, e.g., in communication systems based on such coherence rank (*24*).

Fourth, there are many situations in which the waves from the mutually incoherent sources start out already orthogonal – for example, sparse point sources (e.g., like stars in the sky), widely spaced r.f. antennas, or any other single-mode sources like separate laser beams. Then when we formally write down the "source" coherency matrix $\rho_S$ as in Eq. (3), it is already in its eigenvector and eigenvalue form as in Eq. (7), with the eigenvalues $\mu_{Si}$ corresponding directly to the individual source powers $P_i$ (in non-increasing order). Then, no matter how the resulting field is attenuated and/or scattered, the separable mutually orthogonal received powers, in non-increasing order, are simply bounded by the corresponding original source powers, i.e., $\mu_{Ri} \leq \mu_{Si} = P_i$. This has implications, for example, for r.f. energy harvesting (*26*) from background fields from multiple sources, suggesting an optimal strategy is to design the receiving antenna to concentrate on the strongest source.

Fifth, we can use the Theorem in thought experiments to prove other results in optics. Specifically, we can show that, for arbitrary passive optical systems, for which we can always establish a set of communication mode orthogonal channels (*5*, *10*, *11*), we can prove that we cannot redistribute the sum of their power coupling strengths to generate more orthogonal channels of some weaker coupling. (Supplementary text section S9).

Finally, in Supplementary text section S11 we discuss consequences of the Theorem in common optical situations, resolving some apparent paradoxes.

## Conclusions

We have proved a Maximum Brightness Theorem for radiation through a linear passive optical or wave system. This Theorem applies broadly to waves of many different kinds and to the partially coherent radiation that dominates in natural, thermal, and artificial illumination and many other situations. This Theorem encompasses the classical constant brightness results as well as recent advances (*1*, *2*) (see Supplementary text section S10 for an extended discussion) in one simple statement (Eqs. (1) or (2)); for separable (i.e., mutually orthogonal), mutually incoherent input powers $\mu_{Si}$ and output powers $\mu_{Ri}$, each written in order, we have $\mu_{Ri} \le \mu_{Si}$. What is possibly most surprising or counter-intuitive about this theorem is that it is true regardless of any mixing or scattering of the field as it passes through the passive system; it is not a statement about passing separate "channels" through the system.

We can expect a wide range of applications, consequences and extensions for this very basic Theorem, in optics, r.f. systems, and for waves generally.

**Acknowledgments:** The author is pleased to acknowledge stimulating conversations with Charles Roques-Carmes and Shanhui Fan.

**Funding:** AFOSR Grants FA9550-21-1-0312 and FA9550-23-1-0307

# Supplementary materials

## Methods

### Mathematical proof of the maximum brightness theorem

We note first that, with or without loss, the input and output coherency matrices $\rho_S$ and $\rho_R$ still can be mathematically (and physically (*7*, *8*)) decomposed into mutually incoherent, mutually orthogonal coherent mode components (their eigenfunctions $|\eta_{S_i}\rangle$ and $|\eta_{Ri}\rangle$). As discussed in the main text, we can practically treat such matrices as having finite dimensions, i.e., as $n \times n$ matrices, and the sets $|\eta_{S_i}\rangle$ and $|\eta_{Ri}\rangle$ as having *n* elements. We index these components in non-increasing order of their eigenvalues (and hence of the corresponding powers), i.e., for the eigenvalues $\mu_{Si}$ of the matrix $\rho_S$

$$\mu_{S1} \geq \mu_{S2} \geq \cdots \geq \mu_{Sn} \tag{12}$$

and similarly for the eigenvalues $\mu_{Ri}$ of $\rho_R$

$$\mu_{R1} \geq \mu_{R2} \geq \cdots \geq \mu_{Rn} \tag{13}$$

We characterize the passive optical system with a generally non-unitary operator or matrix $\mathsf{M}$ as in Eq. (5) of the main text. We write this matrix using its singular-value decomposition (*5*) (SVD) as

$$\mathsf{M} = \mathsf{V}\mathsf{D}\mathsf{U}^{\dagger} \tag{14}$$

where $\mathsf{U}$ and $\mathsf{V}$ are unitary and $\mathsf{D}$ is a diagonal matrix with diagonal elements (the singular values) $s_j$, which we can write as

$$\mathsf{D} = \operatorname{diag}\left(s_1, s_2, \ldots, s_n\right) \tag{15}$$

Fig. 2 of the main text shows the concept of a linear optical system in which an input field expressed using the "source" coherency matrix $\rho_S$ is transformed by passing through a passive system represented by $\mathsf{M}$ to generate the received field represented by the "receiving" coherency matrix $\rho_R$. Though the result Eq. (14) might be regarded as just an abstract mathematical operation, in fact an arbitrary optical component can be constructed using this SVD approach, with interferometer meshes implementing unitary matrices and a line of modulators implementing the singular values (*5*, *25*), so we can imagine any such component in at least one implementation. Also shown in Fig. 2 of the main text are the intermediate coherency matrices $\tilde{\rho}_S$ and $\tilde{\rho}_R$ that are important mathematically in our analysis. If the system $\mathsf{M}$ was constructed physically using such an SVD approach (*5*, *25*), these would become actual coherency matrices at these points in the system.

Because this passive system is either lossless or lossy (but has no gain),

$$|s_i|^2 \le 1 \tag{16}$$

Substituting Eq. (14) into Eq. (6) of the main text gives

$$\rho_R = \left(\mathsf{VDU}^\dagger\right)\rho_S\left(\mathsf{UD}^\dagger\mathsf{V}^\dagger\right) \equiv \mathsf{VD}\left(\mathsf{U}^\dagger\rho_S\mathsf{U}\right)\mathsf{D}^\dagger\mathsf{V}^\dagger \tag{17}$$

Now, the matrix

$$\tilde{\rho}_S = \mathsf{U}^\dagger\rho_S\mathsf{U} \tag{18}$$

is just a unitary transformation (by the unitary operator $\mathsf{U}^\dagger$ ) of the matrix $\rho_S$ , so, as shown in Eqs. (7) to (11) of the main text, it will have the same set of eigenvalues, and the matrix

$$\tilde{\rho}_R = \mathsf{V}^\dagger\rho_R\mathsf{V} = \mathsf{V}^\dagger\mathsf{VD}\left(\mathsf{U}^\dagger\rho_S\mathsf{U}\right)\mathsf{D}^\dagger\mathsf{V}^\dagger\mathsf{V} = \mathsf{D}\tilde{\rho}_S\mathsf{D}^\dagger \tag{19}$$

being just a unitary transformation (by the unitary operator $\mathsf{V}^\dagger$ ) of the matrix $\rho_R$ , similarly has the same eigenvalues as the matrix $\rho_R$ . Hence, comparing the eigenvalues of $\rho_R$ and $\rho_S$ is the same as comparing the eigenvalues of $\tilde{\rho}_R$ and $\tilde{\rho}_S$ .

Note too that because the matrix $\rho_S$ is Hermitian (i.e., $\rho_S = \rho_S^\dagger$), so also is the matrix $\tilde{\rho}_S$ ; explicitly,

$$\tilde{\rho}_S^\dagger = \left(\mathsf{U}^\dagger\rho_S\mathsf{U}\right)^\dagger = \left(\mathsf{U}\right)^\dagger\rho_S^\dagger\left(\mathsf{U}^\dagger\right)^\dagger = \mathsf{U}^\dagger\rho_S^\dagger\mathsf{U} = \mathsf{U}^\dagger\rho_S\mathsf{U} = \tilde{\rho}_S \tag{20}$$

Next, we note that, even with complex singular values $s_j$ and hence a complex diagonal matrix $\mathsf{D}$ , the matrix $\tilde{\rho}_R$ is Hermitian, i.e.,

$$\tilde{\rho}_R^\dagger = \left(\mathsf{D}\tilde{\rho}_S\mathsf{D}^\dagger\right)^\dagger = \left(\mathsf{D}^\dagger\right)^\dagger\left(\tilde{\rho}_S\right)^\dagger\mathsf{D}^\dagger = \mathsf{D}\tilde{\rho}_S^\dagger\mathsf{D}^\dagger = \mathsf{D}\tilde{\rho}_S\mathsf{D}^\dagger = \tilde{\rho}_R \tag{21}$$

So this matrix $\tilde{\rho}_R$ also has real eigenvalues. As is generally true for coherency matrices, $\tilde{\rho}_S$ and $\tilde{\rho}_R$ are both positive semi-definite matrices as well as being Hermitian, so the eigenvalues $\mu_{Si}$ and $\mu_{Ri}$ are positive (or zero), and are in non-increasing order as in Eqs. (12) and Eq. (13).

Now we introduce the "$\succeq$" and "$\preceq$" notations: for two positive semi-definite matrices $\mathsf{A}$ and $\mathsf{B}$

$$\mathsf{A} \succeq \mathsf{B} \Leftrightarrow \langle\chi|\mathsf{A}|\chi\rangle \ge \langle\chi|\mathsf{B}|\chi\rangle \tag{22}$$

$$\mathsf{A} \preceq \mathsf{B} \Leftrightarrow \langle\chi|\mathsf{A}|\chi\rangle \le \langle\chi|\mathsf{B}|\chi\rangle \tag{23}$$

for all vectors $|\chi\rangle$ in the Hilbert space of interest. In our case, that Hilbert space is mathematically the set of all *n*-element vectors $|\chi\rangle$ where each element can be an arbitrary complex number (so one could also write $\chi \in \mathbb{C}^n$ ). Formally, we can therefore rewrite the definition of some positive semi-definite matrix $\mathsf{P}$, equivalent to the definition in Eq. (4) in the main text, as

$$\mathsf{P} \succeq 0 \tag{24}$$

(noting an abuse of notation – the "0" on the right is technically an $n \times n$ matrix of zeros, not the number 0). Next, we note that, formally,

$$\mathsf{D}^\dagger\mathsf{D} = \text{diag}\left(|s_1|^2, |s_2|^2, \ldots, |s_n|^2\right) \tag{25}$$

which also means that it is positive semi-definite, as is easily verified using Eq. (4) of the main text. Since $|s_i|^2 \le 1$ (Eq. (16)), then we can formally write

$$\mathsf{D}^\dagger\mathsf{D} \preceq \mathsf{I}_n \tag{26}$$

where $\mathsf{I}_n$ is the $n$-dimensional identity matrix (so, with all $n$ diagonal elements equal to 1).

Using these concepts, we now want to analyze the eigenvalues of $\tilde{\rho}_R = \mathsf{D}\tilde{\rho}_S\mathsf{D}^\dagger$ as in Eq. (19). We note next that, since $\tilde{\rho}_S$ is a positive semi-definite matrix, it has a square root $\tilde{\rho}_S{}^{1/2}$ (see Supplementary text S2) which is also Hermitian (so $\tilde{\rho}_S{}^{1/2} = \left(\tilde{\rho}_S{}^{1/2}\right)^\dagger$) and positive semi-definite, i.e., we can write

$$\tilde{\rho}_S = \tilde{\rho}_S{}^{1/2}\tilde{\rho}_S{}^{1/2} \tag{27}$$

Substituting this into Eq. (19) gives

$$\tilde{\rho}_R = \mathsf{D}\tilde{\rho}_S{}^{1/2}\tilde{\rho}_S{}^{1/2}\mathsf{D}^\dagger \equiv \left(\mathsf{D}\tilde{\rho}_S{}^{1/2}\right)\left(\tilde{\rho}_S{}^{1/2}\mathsf{D}^\dagger\right) \equiv \left(\mathsf{D}\tilde{\rho}_S{}^{1/2}\right)\left(\left(\tilde{\rho}_S{}^{1/2}\right)^\dagger\mathsf{D}^\dagger\right) = \left(\mathsf{D}\tilde{\rho}_S{}^{1/2}\right)\left(\mathsf{D}\tilde{\rho}_S{}^{1/2}\right)^\dagger \tag{28}$$

Now, for any two square matrices $\mathsf{A}$ and $\mathsf{B}$ of the same size, the non-zero eigenvalues of $\mathsf{AB}$ are identical to the non-zero eigenvalues of $\mathsf{BA}$, which can be called the cyclic property of matrix products (see Supplementary text S3). So, we can define a new matrix $\rho'_R$

$$\rho'_R = \left(\mathsf{D}\tilde{\rho}_S{}^{1/2}\right)^\dagger\left(\mathsf{D}\tilde{\rho}_S{}^{1/2}\right) \tag{29}$$

that will have the same eigenvalues $\mu_{Ri}$ as $\tilde{\rho}_R$ and $\rho_R$. Rearranging Eq. (29) gives

$$\rho'_R = \tilde{\rho}_S{}^{1/2}\mathsf{D}^\dagger\mathsf{D}\tilde{\rho}_S{}^{1/2} \tag{30}$$

Because of Eq. (26) and the definition Eq. (23), for any vector $|\xi\rangle$

$$\langle\xi|\mathsf{D}^\dagger\mathsf{D}|\xi\rangle \le \langle\xi|\mathsf{I}_n|\xi\rangle = \langle\xi|\xi\rangle \tag{31}$$

Then we can choose $|\xi\rangle = \tilde{\rho}_S{}^{1/2}|\chi\rangle$, where $|\chi\rangle$ is arbitrary. We therefore have from Eq. (30)

$$\langle\chi|\rho'_R|\chi\rangle = \langle\chi|\tilde{\rho}_S{}^{1/2}\left(\mathsf{D}^\dagger\mathsf{D}\right)\tilde{\rho}_S{}^{1/2}|\chi\rangle = \langle\xi|\mathsf{D}^\dagger\mathsf{D}|\xi\rangle \le \langle\xi|\xi\rangle = \langle\chi|\tilde{\rho}_S{}^{1/2}\tilde{\rho}_S{}^{1/2}|\chi\rangle \tag{32}$$

i.e.,

$$\langle\chi|\rho'_R|\chi\rangle \le \langle\chi|\tilde{\rho}_S|\chi\rangle \tag{33}$$

or equivalently

$$\rho'_R \preceq \tilde{\rho}_S \tag{34}$$

where we note again that $\rho'_R$ and $\tilde{\rho}_S$ have the same eigenvalues $\mu_{Ri}$ and $\mu_{Si}$ as $\rho_R$ and $\rho_S$, respectively. Weyl's monotonicity theorem (see Supplementary text S4) now requires that the ordered eigenvalues $\mu_{Ri}$ and $\mu_{Si}$ of these two matrices satisfy

$$\mu_{R1} \le \mu_{S1},\ \mu_{R2} \le \mu_{S2}, \ldots,\ \mu_{Rn} \le \mu_{Sn} \tag{35}$$

Hence, we have proved the result that, in passing through a passive linear optical system, each eigenvalue of the coherency matrix (and hence the power in each coherency mode), written in non-increasing order, can only stay the same or decrease. Hence, we have proved our Maximum Brightness Theorem, Eq. (1) of the main text.

# Supplementary text

We include below supporting mathematical proofs in sections S1 to S4. These proofs are standard but are included for completeness. The remaining supplementary sections S5 to S11 expand on discussions in the text.

## S1 Transforming a coherency matrix

If we presume that there is a set of possible (normalized) source fields $|\chi_j\rangle$, which are not necessarily orthogonal to one another, and which occur with probabilities $P_j$, then by definition we can write the source density or coherency matrix as

$$\rho_S = \sum_j P_j |\chi_j\rangle\langle\chi_j| \tag{36}$$

For some specific $|\chi_j\rangle$ out of the set of possible fields or sources in the source space, the resulting field after the transformation by this optical system will be

$$|\xi_j\rangle = \mathsf{M}|\chi_j\rangle \tag{37}$$

in the "receiving" space. We can expand $|\xi_j\rangle$ on some complete orthonormal basis $|\beta_u\rangle$ in the receiving space as

$$|\xi_j\rangle = \sum_u b_u^{(j)} |\beta_u\rangle \tag{38}$$

Since $b_u^{(j)} = \langle\beta_u|\xi_j\rangle$ from Eq. (38) and from the orthogonality of the different basis functions $|\beta_i\rangle$

$$\begin{aligned} b_u^{(j)}\left(b_v^{(j)}\right)^* &= \langle\beta_u|\xi_j\rangle\langle\xi_j|\beta_n\rangle \\ &= \langle\beta_u|\mathsf{M}|\chi_j\rangle\langle\chi_j|\mathsf{M}^\dagger|\beta_v\rangle \end{aligned} \tag{39}$$

Quite generally for a coherency matrix $\rho$, it has matrix elements

$$\rho_{mn} = \langle \beta_m | \rho \ | \beta_n \rangle = \sum_j P_j b_m^{(j)} \left( b_n^{(j)} \right)^* \equiv \overline{b_m b_n^*} \equiv \langle b_m b_n^* \rangle \tag{40}$$

where the operation $\langle \cdot \rangle$ corresponds to the ensemble average, as given by the probabilities $P_j$.

So, with a probability $P_j$ of the source state $|\chi_j\rangle$, we have by definition a set of matrix elements

$$\begin{aligned} \langle b_u b_v^* \rangle &= \sum_j P_j \langle \beta_u | \mathsf{M} | \chi_j \rangle \langle \chi_j | \mathsf{M}^\dagger | \beta_v \rangle \\ &= \langle \beta_u | \left( \sum_j P_j \mathsf{M} | \chi_j \rangle \langle \chi_j | \mathsf{M}^\dagger \right) | \beta_v \rangle \end{aligned} \tag{41}$$

The matrix or operator with these matrix elements is, therefore,

$$\rho_R = \sum_j P_j \mathsf{M} | \chi_j \rangle \langle \chi_j | \mathsf{M}^\dagger = \mathsf{M} \left( \sum_j P_j | \chi_j \rangle \langle \chi_j | \right) \mathsf{M}^\dagger \tag{42}$$

i.e.,

$$\rho_R = \mathsf{M} \rho_S \mathsf{M}^\dagger \tag{43}$$

Note this applies whether the matrix $\mathsf{M}$ is unitary or non-unitary.

## S2 Square root of a Hermitian positive semi-definite matrix

Consider a Hermitian positive semi-definite matrix $\mathsf{A}$. Because it is Hermitian, it has real eigenvalues $\lambda_i$, and can be diagonalized by a unitary transformation $\mathsf{Q}$, i.e.,

$$\mathsf{A} = \mathsf{Q} \mathsf{D} \mathsf{Q}^\dagger$$

where $\mathsf{D}$ is a diagonal matrix containing the eigenvalues $\lambda_i$ as its diagonal elements. (The columns of $\mathsf{Q}$ are the eigenvectors.) Because $\mathsf{A}$ is positive semi-definite, the eigenvalues $\lambda_i$ are positive or zero. As a result, we can form a new diagonal matrix that we call $\mathsf{D}^{1/2}$ whose diagonal elements are the positive square roots $\sqrt{\lambda_i}$ of the eigenvalues of $\mathsf{A}$, and for which

$$\mathsf{D}^{1/2} \mathsf{D}^{1/2} = \mathsf{D}$$

Now we propose a matrix

$$\mathsf{B} = \mathsf{Q} \mathsf{D}^{1/2} \mathsf{Q}^\dagger$$

Then we note that

$$\mathsf{B}\mathsf{B} = \mathsf{Q} \mathsf{D}^{1/2} \mathsf{Q}^\dagger \mathsf{Q} \mathsf{D}^{1/2} \mathsf{Q}^\dagger = \mathsf{Q} \mathsf{D}^{1/2} \mathsf{D}^{1/2} \mathsf{Q}^\dagger = \mathsf{Q} \mathsf{D} \mathsf{Q}^\dagger = \mathsf{A}$$

So this matrix $\mathsf{B}$ is indeed a square root of $\mathsf{A}$, which we may choose to write as $\mathsf{A}^{1/2}$.

Note that, because its eigenvalues are real and positive, $\mathsf{B} \equiv \mathsf{A}^{1/2}$ is also a Hermitian positive-semi-definite matrix.

## S3 Cyclic property of matrix products

Consider square matrices $\mathsf{A}$ and $\mathsf{B}$ of the same size. Suppose that $\lambda \neq 0$ is an eigenvalue of the product $\mathsf{AB}$, so with a corresponding eigenvector $|\eta\rangle$. Then

$$(\mathsf{AB})|\eta\rangle = \lambda|\eta\rangle$$

which also shows us that $\mathsf{B}|\eta\rangle \neq \mathbf{0}$ where $\mathbf{0}$ is a vector of zero length. (Otherwise $\mathsf{A}(\mathsf{B}|\eta\rangle) = \mathsf{A}\mathbf{0} = \mathbf{0}$, which implies a zero eigenvalue). Now we multiply both sides by $\mathsf{B}$.

$$\mathsf{B}(\mathsf{AB})|\eta\rangle = \mathsf{B}(\lambda|\eta\rangle)$$

i.e.,

$$(\mathsf{BA})(\mathsf{B}|\eta\rangle) = \lambda(\mathsf{B}|\eta\rangle)$$

Defining $|\kappa\rangle = \mathsf{B}|\eta\rangle$, and noting that we already proved $\mathsf{B}|\eta\rangle \neq \mathbf{0}$, we therefore have an eigenequation for the product $\mathsf{BA}$ with the same (non-zero) eigenvalue $\lambda$ (and associated eigenvector $|\kappa\rangle$. Hence the matrices $\mathsf{BA}$ and $\mathsf{AB}$ have the same non-zero eigenvalues.

## S4 Weyl monotonicity theorem

This theorem can be stated first in the following form:

> Let $\mathsf{A}$ and $\mathsf{B}$ be $n \times n$ Hermitian matrices, with $\mathsf{B} \succeq 0$ (which is the same as saying that $\mathsf{B}$ is positive semi-definite). We order their eigenvalues, written as $\lambda_i(\mathsf{A})$ and $\lambda_i(\mathsf{B})$ respectively, in ascending order, so
>
> $$\lambda_1(\mathsf{A}) \leq \lambda_2(\mathsf{A}) \leq \cdots \leq \lambda_n(\mathsf{A}) \text{ and } \lambda_1(\mathsf{B}) \leq \lambda_2(\mathsf{B}) \leq \cdots \leq \lambda_n(\mathsf{B})$$
>
> Then for every $i = 1, 2, \ldots, n$, we have $\lambda_i(\mathsf{A}) \leq \lambda_i(\mathsf{A}+\mathsf{B})$.

In the proof below, we will be considering a Hilbert space *H* of all *n*-element normalized vectors $|\chi\rangle$ or $|\xi\rangle$ where each element can be an arbitrary complex number (so one could also write $\chi \in \mathbb{C}^n$). We can consider subspaces *K* and $K'$ of *H* (i.e., $K \subseteq H$, $K' \subseteq H$), both of dimensionality $k \leq n$ (i.e., $\dim(K) = \dim(K') = k$). Note that, if a vector $|\chi\rangle$ or $|\xi\rangle$ is in one such subspace (i.e., $|\chi\rangle \in K$, $|\xi\rangle \in K'$), it will still be an *n*-element vector, but the subspaces *K* and $K'$ are each spanned by only *k* orthogonal vectors. The proof relies on the Courant-Fischer minimax theorem, which is the standard theorem underlying the concept of finding eigenvalues and eigenvectors by some variational process. With these definitions, the Courant-Fischer minimax theorem can be stated as follows:

> Let $\mathsf{F}$ be an $n \times n$ Hermitian matrix with eigenvalues $\lambda_1(\mathsf{F}) \leq \lambda_2(\mathsf{F}) \leq \cdots \leq \lambda_n(\mathsf{F})$. ($\mathsf{F}$ therefore operates on the Hilbert space *H*.) Then the *k*th eigenvalue of $\mathsf{F}$ is given by

$$\lambda_k(\mathsf{F}) = \min_K \max_{|\chi\rangle \in K} \langle \chi | \mathsf{F} | \chi \rangle$$

(The $|\chi\rangle$ established in this process will be the (normalized) eigenvector corresponding to this eigenvalue.)

We will not prove the Courant-Fischer minimax theorem, but it can be useful to understand the process it represents. We choose some *k*-dimensional subspace *K*. We work through all the (normalized) vectors in that space to maximize $\langle \chi | \mathsf{F} | \chi \rangle$. Then we choose another such subspace *K*, repeating the process of going through all the vectors in that subspace. We keep doing this until we find the minimum possible value of this $\langle \chi | \mathsf{F} | \chi \rangle$ for any choice of such a *k*-dimensional subspace. The reason we use this minimization is to eliminate from the subspace the vectors that would correspond to larger eigenvalues. The maximization within that subspace has then found the largest possible eigenvalue when all those other vectors corresponding to larger eigenvalues have been excluded.

We now return to the proof of Weyl's monotonicity theorem.

We note that, because of linearity, for any vector $|\chi\rangle$ in *H*,

$$\langle \chi | \mathsf{A} + \mathsf{B} | \chi \rangle = \langle \chi | \mathsf{A} | \chi \rangle + \langle \chi | \mathsf{B} | \chi \rangle \tag{44}$$

Since $\mathsf{B}$ is positive semi-definite, by definition $\langle \chi | \mathsf{B} | \chi \rangle \geq 0$ for any vector $|\chi\rangle$ in $H$, so we see that on the right we are always adding a non-negative number $\langle \chi | \mathsf{B} | \chi \rangle$ to the number $\langle \chi | \mathsf{A} | \chi \rangle$. We need to consider the two minimizations

$$\lambda_k(\mathsf{A} + \mathsf{B}) = \min_{K'} \max_{|\xi\rangle \in K'} \langle \xi | \mathsf{A} + \mathsf{B} | \xi \rangle \tag{45}$$

and

$$\lambda_k(\mathsf{A}) = \min_K \max_{|\chi\rangle \in K} \langle \chi | \mathsf{A} | \chi \rangle \tag{46}$$

where we are being careful not to presume the same spaces and vectors in the two different expressions Eqs. (45) and (46). We note that the minimization in Eq. (45) can never lead to a smaller result than $\min_{K'} \max_{|\xi\rangle \in K'} \langle \xi | \mathsf{A} | \xi \rangle$, because we are always adding a non-negative number $\langle \xi | \mathsf{B} | \xi \rangle \geq 0$ in Eq. (45). So

$$\lambda_k(\mathsf{A} + \mathsf{B}) \geq \min_{K'} \max_{|\xi\rangle \in K'} \langle \xi | \mathsf{A} | \xi \rangle$$

But

$$\min_{K'} \max_{|\xi\rangle \in K'} \langle \xi | \mathsf{A} | \xi \rangle \equiv \min_K \max_{|\chi\rangle \in K} \langle \chi | \mathsf{A} | \chi \rangle = \lambda_k(\mathsf{A}) \tag{47}$$

since there is only a notational difference between these expressions. So, finally,

$$\lambda_k(\mathsf{A} + \mathsf{B}) \geq \lambda_k(\mathsf{A}) \tag{48}$$

Having proved Weyl's monotonicity theorem, we can now state it in a form that is more convenient for our case. Consider an $n \times n$ Hermitian matrix $\mathsf{C} \succeq \mathsf{A}$. We can then define a positive semi-definite matrix $\mathsf{B} = \mathsf{C} - \mathsf{A} \succeq 0$, and write $\mathsf{A} + \mathsf{B} = \mathsf{C}$. Substituting into Eq. (48), we can therefore state an alternative form of Weyl's monotonicity theorem:

> Let $\mathsf{C}$ and $\mathsf{A}$ be $n \times n$ Hermitian matrices with $\mathsf{C} \succeq \mathsf{A}$. We order their eigenvalues, written as $\lambda_i(\mathsf{A})$ and $\lambda_i(\mathsf{C})$ respectively, in ascending order, so
>
> $$\lambda_1(\mathsf{A}) \le \lambda_2(\mathsf{A}) \le \cdots \le \lambda_n(\mathsf{A}) \text{ and } \lambda_1(\mathsf{C}) \le \lambda_2(\mathsf{C}) \le \cdots \le \lambda_n(\mathsf{C})$$
>
> Then for every $i = 1, 2, \ldots, n$, we have $\lambda_i(\mathsf{A}) \le \lambda_i(\mathsf{C})$.

Finally, for our purposes it will be more convenient to number eigenvalues in non-increasing order. This just requires a notational change in the indices. So, using the corresponding eigenvalues as written in Eq. (12) and Eq. (13), we change notation to

$$\mu_{S1} = \lambda_n(\mathsf{C}),\ \mu_{S2} = \lambda_{n-1}(\mathsf{C}),\ \ldots,\ \mu_{S(n-1)} = \lambda_2(\mathsf{C}),\ \mu_{Sn} = \lambda_1(\mathsf{C})$$

$$\mu_{R1} = \lambda_n(\mathsf{A}),\ \mu_{R2} = \lambda_{n-1}(\mathsf{A}),\ \ldots,\ \mu_{R(n-1)} = \lambda_2(\mathsf{A}),\ \mu_{Rn} = \lambda_1(\mathsf{A})$$

So, choosing the matrices $\mathsf{A} \equiv \rho'$ and $\mathsf{C} \equiv \tilde{\rho}_S$ with $\rho' \preceq \tilde{\rho}_S$ as in Eq. (34) or equivalently $\rho_S \succeq \rho'$, Weyl's monotonicity theorem tells us that for all $i = 1, 2, \cdots n$, $\mu_{Si} \ge \mu_{Ri}$, which is the core result as shown in Eq. (35) and Eq. (1) of the main text.

## S5 Majorization bound on eigenvalues

The majorization bound on the coherency matrix eigenvalues is introduced in Ref. (*2*). The topic of majorization has recently been of growing interest in optics; see also (*27–32*). Suppose we take two sets **a** and **b**, each of *n* numbers, and write them in decreasing (or more rigorously non-increasing) order, so lists notated as $a_1^{\downarrow}$ and $b_i^{\downarrow}$. Then, **a** majorizes **b**, written as $a \succ b$, if $\sum_{i=1}^{k} a_i^{\downarrow} \ge \sum_{i=1}^{k} b_i^{\downarrow}$ for all *k* from 1 to *n*. For strict majorization, we also require $\sum_{i=1}^{n} a_i^{\downarrow} = \sum_{i=1}^{n} b_i^{\downarrow}$; with loss in the system, this second condition will not be satisfied for the input eigenvalues majorizing the output eigenvalues, in which case we can more rigorously state that the input eigenvalues weakly majorize the output eigenvalues (notated using $\succ_w$).

First, we note that the results of the Maximum Brightness Theorem are consistent with the majorization bound – that is, if we satisfy the Maximum Brightness Theorem, then the output eigenvalues are indeed (weakly) majorized by the input eigenvalues. The proof of this is straightforward. For each input eigenvalue $\mu_{Si}$ and the corresponding output eigenvalue $\mu_{Ri}$ (with both lists of eigenvalues written in non-increasing order), the Maximum Brightness Theorem requires $\mu_{Si} \ge \mu_{Ri}$. So, we could write $\mu_{Si} = \mu_{Ri} + \varepsilon_i$ for some number $\varepsilon_i \ge 0$. So, $\sum_{i=1}^{k} \mu_{Si} = \sum_{i=1}^{k} \mu_{Ri} + \sum_{i=1}^{k} \varepsilon_i \ge \sum_{i=1}^{k} \mu_{Ri}$, which holds for all *k*. Hence, the Maximum Brightness Theorem does indeed require that the output eigenvalues are (weakly) majorized by the input eigenvalues, as proposed in Ref. (*2*), and so that work is quite correct in its conclusion that the output eigenvalues are majorized by the input eigenvalues (technically, as a weak majorization).

However, the Maximum Brightness Theorem introduced here, as well as being a statement about each pair of input and output eigenvalues, is also a stronger statement. Specifically, we can find sets of numbers that satisfy majorization but do not satisfy the Maximum Brightness Theorem. Consider, for example, an input coherency matrix with a set of input eigenvalues as the list, in decreasing order, $\boldsymbol{\mu}_S = [\mu_{S1} = 1,\ \mu_{S2} = 0.3]$. Presume we propose a corresponding list of eigenvalues of the output coherency matrix, in decreasing order, $\boldsymbol{\mu}_R = [\mu_{R1} = 0.8,\ \mu_{R2} = 0.4]$. Then we can see that $\boldsymbol{\mu}_S$ weakly majorizes $\boldsymbol{\mu}_R$, i.e., $\boldsymbol{\mu}_S \succ_w \boldsymbol{\mu}_R$. Specifically,

$$\sum_{i=1}^{1} \mu_{Si} = 1 \geq \sum_{i=1}^{1} \mu_{Ri} = 0.8$$

$$\sum_{i=1}^{2} \mu_{Si} = 1 + 0.3 = 1.3 \geq \sum_{i=1}^{2} \mu_{Ri} = 0.8 + 0.4 = 1.2$$

This output set of eigenvalues would be allowed if we only required such majorization. However, it is not allowed from our Maximum Brightness Theorem because $\mu_{R2} = 0.4 > \mu_{S2} = 0.3$, i.e., the second largest output power is greater than the second largest input power. Hence, the Maximum Brightness Theorem, while including the majorization result, is a more restrictive bound.

## S6 Scattering matrix and non-reciprocity

Because the constant brightness theorem is usually discussed in terms of light transmitted through some optical system, we introduced the optical system and its associated operator $\mathsf{M}$ in this "transmission matrix" way rather than as a scattering matrix $\mathsf{S}$. The core difference with a scattering matrix is that we explicitly allow the backwards versions of the input channels also to be output channels and the backwards versions of the output channels also to be input channels. The scattering matrix simply relates all the "outgoing" waves to all the "ingoing" waves.

In fact, in our discussion, though for descriptive simplicity we have shown inputs on the left mapping to outputs on the right, the input channels can be coming from any direction and the output channels can be going in any directions (including backwards versions of the input channels); we have not had to be restrictive about what the input and output channels are, so our analysis includes these possibilities for the input and output channels. Furthermore, at no point have we restricted to reciprocal optics. So, the results derived here apply generally to the scattering matrix also, and to non-reciprocal optics.

## S7 Maximum channel brightness and total brightness

We can ask what is the maximum possible separated output power for some given input coherency matrix $\rho_S$ with eigenvalues $\mu_{Si}$ in non-increasing order after we have passed through some passive optical system $\mathsf{M}$. If the optical system $\mathsf{M}$ has input communication mode functions $|\psi_i\rangle$ and singular values $s_i$, all indexed in non-increasing order of the singular value magnitudes, then the obvious choice is if the first coherency matrix eigenfunction (coherent mode) $|\eta_{S1}\rangle$ happens to be the same as the input communication mode function $|\psi_1\rangle$. So, the best possible coupling would be if $|\eta_{S1}\rangle = |\psi_1\rangle$, with corresponding output coherent mode power or **Maximum Channel Brightness** of

$$P_1 = \mu_{R1} = |s_1|^2 \mu_{S1} \tag{49}$$

Suppose now we want to get the largest possible total power though the system. Then the best we can do is to continue to have the eigenvectors $|\eta_{Si}\rangle$ generally correspond to the communication mode input vectors $|\psi_i\rangle$ of $\mathsf{M}$. Then with $P_1$ as above, the next largest power component we can have is similarly if $|\eta_{S2}\rangle = |\psi_2\rangle$, and so on, leading to a total power $P_{tot}$ and a **Maximum Total Brightness Theorem**

$$P_{tot} \le \sum_{i=1}^{n} |s_i|^2 \mu_{Si} \tag{50}$$

(Note, incidentally that any rearrangement of the coherent modes among the input communication mode functions – e.g., associating $|\psi_2\rangle$ with $|\eta_{S1}\rangle$ and $|\psi_1\rangle$ with $|\eta_{S2}\rangle$ – only leads to a smaller total, consistent with the inequality in Eq. (50) and as supported mathematically by the "rearrangement inequality", several proofs of which are available.)

Note that, if we start with a specific such input coherency matrix $\rho_S$ whose eigenvectors do not correspond to the input communication mode vectors $|\psi_i\rangle$ of the system $\mathsf{M}$, we could introduce a unitary interferometer mesh to implement the necessary unitary transform to change to those communication mode vectors before entering the system $\mathsf{M}$. That unitary transform would be, formally

$$\mathsf{U}_S = \sum_{i=1}^{n} |\psi_i\rangle\langle\eta_{Si}| \tag{51}$$

Since this unitary transformation is the one that will give the maximum power as above, we could find it by inserting a programmable unitary transformer (e.g., an interferometer mesh) at the input to the optical system and performing a global total power maximization over this unitary transform. We would end up connecting the input coherent modes one by one to the corresponding output coherent modes, with the eigenvalues of the output coherency matrix $\rho_R$ (and the actual powers in those output eigenvectors) $|\eta_{Ri}\rangle$) of $\mu_{Ri} = |s_i|^2 \mu_{Si}$, similar to the approach in Ref. (*14*). We should also be able to do this power optimization channel by channel, starting with the most powerful by globally optimizing self-configuring layers of the mesh one by one, as in Refs. (*7*, *8*). Note this leads to actual orthogonal channels from inputs to outputs associated with mapping input coherent mode $|\eta_{Si}\rangle$ to output coherent mode $|\eta_{Ri}\rangle$.

## S8 Bounds on rank

If the input coherency matrix $\rho_S$ has a finite rank *r*, then, as discussed in (*1*, *24*) and in a thermodynamic proof of conservation of lossless channels in (*12*), this rank (called wave étendue in (*1*)) is conserved in lossless systems. If we know that the field is generated by some specific number $r$ of mutually incoherent sources, then that rank is indeed finite at $r$. However, for a propagating partially coherent light field generated, e.g., from natural sources, it may not be simple to write down a definite rank. Generally, the communication mode approach will lead to a formally infinite number of channels, even though, as argued above, we can practically truncate it to some number $n$ (e.g., at or somewhat above the tunneling escape limit (*19*)). In general, too, because of these constraints on propagating channels, for the output or received coherency matrix $\rho_R$, though the powers may be moderately large up to some number and substantially weaker thereafter, there may formally be no definite rank.

We can address these issues by defining what we can call a threshold rank $r_\varepsilon$. Specifically, we choose some minimum threshold power value $\varepsilon$ at or below which we regard any subsequent coherency matrix eigenvalues to be negligible. Then the threshold rank $r_\varepsilon$ is the number of input eigenvalues $\mu_{Si}$ greater than $\varepsilon$, i.e., for which $\mu_{Si} > \varepsilon$.

The Maximum Brightness Theorem certainly requires that we cannot add another $(r+1)$th eigenvalue to the output coherency matrix for which $\mu_{R(r+1)} > \varepsilon$. That would require $\mu_{S(r+1)} > \varepsilon$, contradicting the assumption of only $r_\varepsilon$ input eigenvalues $\mu_{Si} > \varepsilon$. In a lossless system, the eigenvalues are not changed. So, we can state the following:

> **Threshold rank bound:** the threshold rank of the coherency matrix (the number of coherent modes of power greater than some threshold $\varepsilon$) cannot increase in a passive linear optical system and is conserved in a lossless system.

Note that, though this threshold rank cannot increase, it is quite possible that it can decrease in a lossy system. For example, the singular values could fall off so that the output powers of at least some of the separable components became below some chosen detection threshold, hence reducing the threshold rank.

Recent work (*24*) shows that, if we construct an input coherency matrix with finite rank, then in a lossless system or one with uniform loss over all relevant channels (constant power coupling strengths $|s_i|^2$ over sufficient channels), the rank of that matrix can be used to communicate information despite strong scattering, consistent with the "wave étendue" or rank conservation; each separated output power could be a combination of any or all of the input powers, but the rank would be retained given a sufficiently low detection power threshold. We cannot guarantee that this rank or threshold rank will be retained in lossy systems. However, by employing some protocol to test periodically that the full range of ranks was still measurable at the receiving end, this approach (*24*) could likely still be used in a general lossy system.

## S9 Proof that we cannot increase the number of moderately coupled channels

Suppose we have some optical system with a source volume $V_S$, a receiving volume $V_R$, and a coupling operator $G_{SR}$ between them, as in the communication mode analysis in, e.g., Refs. (*5*,

*10*). This coupling operator can represent any passive optical system – it need not just be some free-space Green's function. Because we believe that the physics of waves is such that finite sources in $V_S$ give finite waves in $V_R$, we can construct a sum rule involving the amplitude connection strengths $g_{ij}$ from the *j*th member of some set of (normalized) source functions in $V_S$ and the *i*th member of some set of (normalized) wave functions in $V_R$ (*5*, *10*). Before analyzing what are the actual orthogonal (i.e., zero crosstalk) channels we might construct for this system, we can evaluate the quantity *S* (the square of the Hilbert-Schmidt norm) which corresponds to the sum of the squares of the amplitude connection strengths. For example, in the simple scalar wave case, $\mathsf{G}_{SR}$ at some angular frequency would correspond to some scalar Green's function $G_\omega(\mathbf{r}_R;\mathbf{r}_S)$ relating amplitudes at source points $\mathbf{r}_S$ to the resulting amplitudes at receiving points $\mathbf{r}_R$, and we could write

$$S = \int_{V_R}\int_{V_S} \left|G_\omega(\mathbf{r}_R;\mathbf{r}_S)\right|^2 d^3\mathbf{r}_S d^3\mathbf{r}_R = \sum_{i,j}\left|g_{ij}\right|^2 \tag{52}$$

Similar formulas exist for other forms of waves, such as vector waves with dyadic Green's functions (*5*).

We might hope that we might be able to choose our orthogonal channels in any way that satisfies this sum rule. We might want a large number of moderately well coupled channels, for example, rather than a small number of well coupled channels. However, once we are given the Green's function or, equivalently, the coupling operator $\mathsf{G}_{SR}$ for the system, we will find we do not have that freedom, and we can prove this and other useful results using a thought-experiment and the Maximum Brightness Theorem. Specifically, we will be able to prove that we cannot construct any passive linear optical apparatus that allows us to perform this desired reapportionment of the coupling strength to different, larger numbers of modes.

First, we rewrite $\mathsf{G}_{SR}$ using its singular value decomposition (SVD) as

$$\mathsf{G}_{SR} = \mathsf{V}\mathsf{D}\mathsf{U}^\dagger \tag{53}$$

where $\mathsf{U}$ and $\mathsf{V}$ are unitary matrices and the diagonal matrix $\mathsf{D} = \mathrm{diag}(s_{G1}, s_{G2}, \ldots s_{Gn})$ containing the singular values $s_{Gk}$ as its diagonal elements, which we can choose order in non-increasing order of their magnitude, and *n* is a large enough integer. Incidentally, if we choose to continue possibly to an infinite value of *n*, we would also have (*5*, *10*)

$$S = \sum_{k=1}^{\infty}\left|s_{Gk}\right|^2 \tag{54}$$

So, in practice, we can truncate the sum to some finite *n* once we have approached this limit *S* to any sufficient degree.

Already we can suspect that we will never do any better than the SVD channels through the optical system – i.e., the communication modes – with their coupling strengths given by the singular values, but we can prove some specific results here.

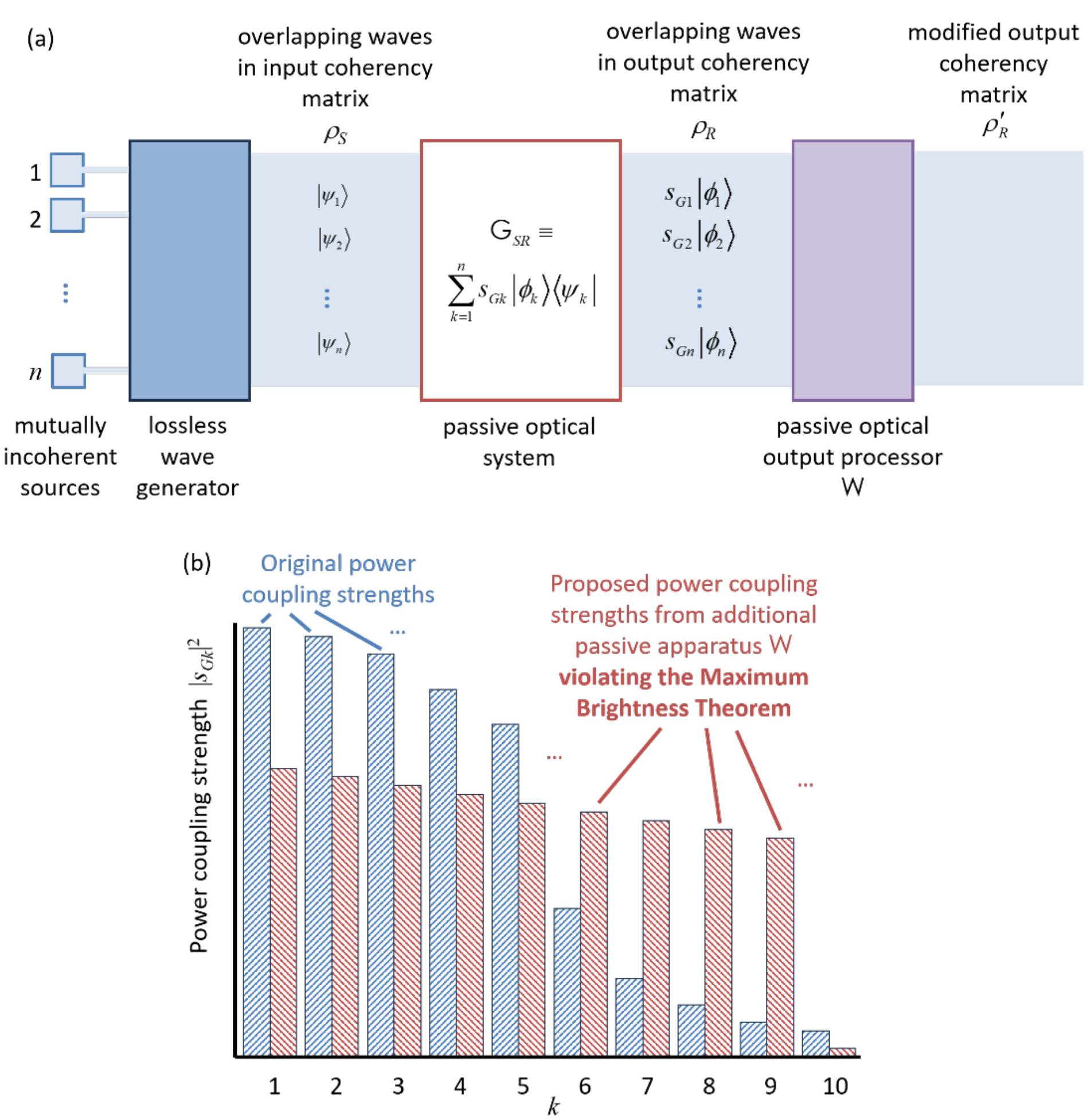

Fig. S1. Thought experiment proving power coupling strengths cannot be redistributed. (a) Hypothetical apparatus to try to redistribute the power coupling strengths $|s_{Gk}|^2$ of a passive optical system described by coupling operator $G_{SR}$, written here in the SVD form mapping orthogonal input vectors $|\psi_k\rangle$ to orthogonal output vectors $|\phi_k\rangle$. (b) The hypothetical passive optical output processor would try to reapportion the coupling strengths while retaining their overall sum, as shown in the proposed power coupling strengths in (b). But, such an apparatus would lead to violation of the Maximum Brightness theorem because the resulting hypothetical modified output coherency matrix would correspond for k = 6, 7, 8, or 9 to increasing the eigenvalue (and power) of those modes.

Imagine the thought-experiment apparatus of Fig. S1. We start with $n$ mutually incoherent sources, all of unit power. We then use these to form, one by one, waves with the spatial form of the corresponding communication mode input or source functions $|\psi_k\rangle$ associated with the coupling operator $G_{SR}$. The resulting set of waves is mutually orthogonal and mutually incoherent, so expresses the input coherency matrix $\rho_S$. Each of these will then be coupled through $G_{SR}$ to give an orthogonal output wave, of normalized form $|\phi_k\rangle$, in the receiving space, with coupling amplitude $s_{Gk}$. These output waves are all mutually orthogonal (because the $|\phi_k\rangle$ are mutually orthogonal) and mutually incoherent (because each one is formed from the light from exactly one of the original mutually incoherent sources). Hence, these $|\phi_k\rangle$ form the eigenvectors of the output coherency matrix $\rho_R$, with associated eigenvalues $|s_{Gk}|^2$.

Suppose then we try to construct some linear passive optical system $\mathsf{W}$ to act on the wave in the receiving space to reapportion the power coupling strengths $|s_{Gk}|^2$ in some way over the orthogonal output channels, which would create some new output coherency matrix $\rho'_R$. That will correspond to trying to reapportion the eigenvalues of the coherency matrix and specifically trying to increase the power in some of the higher-numbered, weaker coherent modes or output channels of this new output coherency matrix $\rho'_R$. Then we know from the Maximum Brightness Theorem that we cannot do this. All we can do is to make any resulting eigenvalues (and hence power coupling strengths) possibly weaker. Specifically, we cannot take a small number of strong coupling strengths and reapportion them among a larger number of possible orthogonal channels because that would correspond to increasing the eigenvalues of the coherency matrix in those other orthogonal output channels (i.e., coherent modes). If we could make a machine that would have redistributed the singular values squared in this way, we would have violated the Maximum Brightness Theorem.

## S10 Relation to brightness and étendue

Here we will summarize the concept of brightness in classical optics and how it relates to the discussion in this paper. We try to make this discussion relatively self-contained for greater readability.

In conventional or classical discussions in the general field of radiometry (*3*), radiance (also known as brightness) is the power per unit projected area per unit solid angle, and is a property of the light field. (By projected area we mean that, if we are looking primarily at some viewing angle $\theta$ relative to the normal to the surface area $A$ of some source, the projected area of that source is $A\cos\theta$.) Suppose that the entrance pupil of the optics effectively subtends the same solid angle $\Omega$ for all parts of the source (which will be effectively true if the entrance pupil is much larger than the source area). Presuming that emission from any unit projected area into each small element of solid angle is equal (a so-called Lambertian source), then the total power captured by the optics in this picture is just the integral of the radiance over the surface area $A$ of the emitter and the solid angle of the entrance pupil

$$P_{TOT} = \int_{\substack{\text{source}\\ \text{area}}} \int_{\substack{\text{solid}\\ \text{angle}}} L(\mathbf{r})\, dA \cos\theta\, d\Omega \tag{55}$$

If we also presume that the source is uniform, with therefore the same radiance $L_o$ at all points, then we can write this as

$$P_{TOT} = L_o \int_{\substack{\text{source}\\ \text{area}}} \int_{\substack{\text{solid}\\ \text{angle}}} L(\mathbf{r})\, dA \cos\theta d\Omega \tag{56}$$

which we can rewrite as

$$P_{TOT} = L_o \mathcal{E} \tag{57}$$

where (presuming unit refractive index in our definition for simplicity) $\mathcal{E}$ is the quantity called étendue, given by

$$\mathcal{E} = \int_{\substack{\text{source}\\ \text{area}}} \int_{\substack{\text{solid}\\ \text{angle}}} dA \cos\theta d\Omega \tag{58}$$

(Conventionally, with a refractive index $n_o$, we would introduce a factor $1/n_o^2$ on the right of Eq. (57) and $n_o^2$ on the right of Eq. (58).) The étendue is a property of the optical system, not of the optical field. If the system is lossless, then, because the power is conserved, so also must étendue be conserved. In a ray picture, the conservation of étendue is essentially the same as saying that a lossless optical system retains all rays as they progress through the system.

This classical picture with radiance and étendue is very useful for large, incoherent optical systems with Lambertian emitters. However, in this simple form it does not cover other forms of emitters with different angular behavior, it does not include wave effects such as diffraction, and so is not suitable once we approach wavelengths scales.

Recently, modal pictures of light and waves have become increasingly well developed. Such modal pictures can implicitly include all diffraction effects. The communication mode picture is very useful for considering complete wave systems, from sources to receivers (*5*, *10*). It defines a mutually orthogonal set of source functions $|\psi_i\rangle$ that couple one by one to a mutually orthogonal set of waves $|\phi_i\rangle$ in some receiving space, with amplitude coupling strengths given by the singular values $s_i$ of the coupling operator $G_{SR}$. These functions and singular values are deduced directly from the singular value decomposition of the coupling operator $G_{SR}$.

If we had just some finite number *m* of non-zero singular values $s_i$ (so the coupling operator $G_{SR}$ has rank *m*), then we could define a quantity $\mathcal{E}_m$ given by a sum over those singular values rather than the integral of Eq. (58), giving

$$\mathcal{E}_m = \sum_{i=1}^{m} |s_i|^2 \tag{59}$$

If the power in every one of the *m* input source functions $|\psi_i\rangle$ was the same value $P_o$, then the total output power would be

$$P_{TOT} = P_o \mathcal{E}_m \tag{60}$$

by analogy with Eq. (57) above. If the system was lossless, so every $s_i = 1$, then $\mathcal{E}_m = m$, the rank of the coupling operator. Recently, this rank has indeed been proposed as a useful concept called the “modal étendue” (*1*). We note, too, that having a specific number of singular values all the same or similar is quite a common situation especially in paraxial optics, where we can refer to it as “paraxial degeneracy” (*5*); this is one reason why such approaches work in classical optics. This result Eq. (60) also serves as a constant brightness theorem for a lossless optical system in this modal picture, encompassing the previous classical result.

In the modal picture, we can usefully go beyond this description, however. We can first be much more specific about sources, not just presuming the sources are Lambertian. Imagine that we have a set of $q$ different emitters, all mutually incoherent. These emitters could even be individual atoms, each of which emits a specific form of wave $|\chi_i\rangle$. As in Eq. (3) in the main text, we write this as a coherency matrix $\rho_S = \sum_{i=1}^{q} P_i |\chi_i\rangle\langle\chi_i|$ . The $P_i$ here can be interpreted as the average power from such emitters, but we could instead interpret it as the probability per unit time that this particular emitter or “atom” $i$ emits a photon into this field. Note, to emphasize, that these waves $|\chi_i\rangle$ need not be orthogonal to one another; for example, if the emitters or “atoms” are very close to one another, for example, much less than half a wavelength, these $|\chi\rangle$ will almost certainly not be orthogonal.

We then formally find the eigenvectors $|\eta_{Si}\rangle$ (coherent modes) and eigenvalues $\mu_{Si}$ of this coherency matrix. These eigenvectors $|\eta_{Si}\rangle$ are orthogonal to one another, and the $\mu_{Si}$ now represent the powers in these eigenvector waves. We note that we can build a physical machine (*7*, *8*) that measures and separates the field into exactly these eigenvectors, with the powers $\mu_{Si}$ appearing at its individual outputs, so these phenomena are quite real and measurable.

As a practical matter, we will only need the first $n$ of these, with $n$ possible much less than $q$, because quite generally in propagation the number of such orthogonal modes that can propagate will have some relatively abrupt cut-off in power coupling strengths (*5*, *19*).

Now for our source, we can describe it using these $n$ orthogonal source functions $|\eta_{Si}\rangle$, each also mutually incoherent with every other such source function, which also means we can simply add the powers in such sources to get the total power in the wave.

We can think of the $\mu_{Si}$ as being the “brightness” of each of these “modes”. Note now we are discussing “brightness” as the power, or photon probability per unit time, of a mode, not the power per unit solid angle per unit area, so we could call each $\mu_{Si}$ a “modal brightness”.

So, our initial total power, rather than being an integral over area and solid angle, is now a sum over modes

$$P_{TOT} = \sum_{i=1}^{q} \mu_{Si} \cong \sum_{i=1}^{n} \mu_{Si} \tag{61}$$

Now we pass this into our optical system $G_{SR}$, with input communication modes $|\psi_j\rangle$. This set $|\psi_j\rangle$ can be complete for the input wave space; if not, we can always extend it (possibly using further orthogonal functions corresponding to zero singular values). So, we can express any input eigenfunction $|\eta_{Si}\rangle$ as an expansion in the $|\psi_j\rangle$. If the $|\eta_{Si}\rangle$ can all be completely expressed in terms $|\psi_j\rangle$ corresponding to unit (i.e., lossless) singular values $s_i$, then, writing the input and output eigenvalues $\mu_{Si}$ and $\mu_{Ri}$ each in lists that are decreasing (non-increasing), the eigenvalues of the output coherency matrix will be equal to the input eigenvalues ($\mu_{Ri} = \mu_{Si}$), which is a (modal) constant brightness result that also conserves overall power (so a "total" constant brightness result). For completeness, we can restate this lossless result (which is already known (*1*, *24*)), as a special case of the Maximum Brightness theorem:

> **(Modal) Constant Brightness Theorem:** In passing through a lossless optical system, the power in the *i*th most powerful mutually incoherent, mutually orthogonal component of the field remains the same.

Obviously, if the powers in each coherent mode are preserved and all coherent modes are communicated losslessly by the system, the total power is conserved, which we could call a **Constant Total Brightness Theorem** for lossless systems, though this is immediately obvious from the fact that the system is lossless.

If there is loss in the system, so for some singular values $|s_i|^2 < 1$ and/or if we cannot express the $|\eta_{Si}\rangle$ completely in terms of the $|\psi_j\rangle$, then from the Maximum Brightness Theorem we know that $\mu_{Ri} \le \mu_{Si}$, which is a modal maximum brightness result. Then, the total output power

$$P_{OUT} = \sum_{i=1}^{q} \mu_{Ri} \le \sum_{i=1}^{q} \mu_{Si} \tag{62}$$

which is the Maximum Total Brightness theorem above in section S7.

We note too that Supplementary Material Section 1 of Ref. (*1*) provides an extended justification of why, in the lossless case, the "wave étendue" bound, or equivalently the conservation of the rank of the coherency matrix, is equivalent to the classical constant brightness theorem. Also, as discussed above, the Maximum Brightness Theorem encompasses the majorization bound (*2*) (see Supplementary text section S5), and is tighter because some situations that would be allowed if majorization was the only criterion are forbidden by the Maximum Brightness Theorem. Ref. (*2*) has argued that majorization of the eigenvalues is sufficient to guarantee that the brightness has not increased. Since our result is also consistent with majorization, so also our approach guarantees that brightness has not increased in this view.

So, the Maximum Total Brightness theorem can be viewed as encompassing the previous classical constant brightness theorem result as a special case. It allows more specific results fully including diffraction effects and allowing for different brightnesses (powers) of different modes in the system. It also encompasses recent results (*1*, *2*), with bounds that are tighter in some cases.

## S11 Consequences of the Maximum Brightness Theorem in common optical situations

In this section, we respond to a number of simple questions one could ask about the consequences of this Theorem in a number of situations, resolving some potential paradoxes.

Can I take the power in a small number of mutually incoherent sources and spread them out to a large number of different separate receivers without violating the Theorem?

> Yes, but if we took the power arriving at, say, some pair of such receivers and routed those powers to an interferometer, we would in general see interference fringes, at least of some depth. Though the different receivers, being separate, can be regarded as being orthogonal, these pairs of powers would not generally be mutually incoherent. The theorem makes statements about components of the field that are both mutually orthogonal and mutually incoherent.

Can I take the power in a large number of mutually incoherent mutually orthogonal sources and have power from each of these sources arriving at a smaller number of receivers without violating the Theorem?

> Yes, but the theorem tells us (among other things) that the power at any one of those receivers can never exceed that of the brightest source, and also that we could not concentrate the total power into that smaller number of receivers.

I have a large number of atoms, each of which emits independently from all the others, so they constitute mutually incoherent sources. Does the Theorem mean that the largest power I can receive is the power from just one atom?

> No. Unless they are spaced far enough apart, the light from the different individual atoms is not into mutually orthogonal beams. Analyzing the situation of dense or continuous sets of mutually incoherent sources has been a minor formal problem in the theory of partial coherence, one that has been resolved with some *ad hoc* assumptions, such as neglecting evanescent waves (*33*). This has led to a "grouping" into effective areas of $\sim \lambda^2 / \pi$, which will generally have a larger power than that just from one atom (e.g., in some thermal emitter). We will not analyze this further here, however, though we note that the recent tunneling escape analysis of waves (*19*) also rigorously deduces a $\lambda^2 / \pi$ effective area per well-coupled mode without resorting to simple neglect of evanescent waves.
>
> If the atoms are far enough apart (e.g., by distance much larger than a wavelength), the power in the strongest coherent mode is bounded by the maximum total power emitted by one atom, however, no matter how that light is subsequently scattered or absorbed.